\documentclass[pss]{wiley2sp} % provides pss two-column style
\usepackage{amsmath}
\begin{document}

% Title of the article
\title{Radiation fields for nanoscale systems}

% Abbreviated title for the page headers
\titlerunning{Short title }

% Authors
\author{%
 MingLiang Zhang,
  D.A. Drabold\textsuperscript{\Ast},
  }

% Abbreviated list of authors for the page headers
\authorrunning{First author et al.}

%E-mail-address of corresponding author
\mail{e-mail
  \textsf{zhangm@ohio.edu,~~drabold@ohio.edu}
 }

% author's affiliations/addresses
\institute{%
  Department of Physics and Astronomy, Ohio University, Athens, Ohio 45701
  }

\received{XXXX, revised XXXX, accepted XXXX} % do not change, will be filled in by the publisher
\published{XXXX} % do not change, will be filled in by the publisher

% Please select about four verbal keywords for your manuscript.
\abstract{%
% This is a macro for the typesetting of two-column text in an
% abstract. It will typeset the two arguments in \abstcol{}{} as the
% left and right column inside the abstract box. At the
% columnbreak there will be always a columnbreak (\par), so both
% columns start with a new paragraph. No automatic column height
% balancing is done.
%
% If used with a \titlefigure it will silently output both
% parameters as consecutive paragraphs.
%
% The macro is defined exclusively inside the argument of \abstract{};
% if used outside it will raise an error.
%
% Usage: \abstcol{<left column>}{<right column>}
\abstcol{%
For a group of charged particles obeying quantum mechanics interacting with
an electromagnetic field, the charge and current density in a pure
state of the system are expressed with the many-body wave function of the state. Using these as sources,
 the microscopic Maxwell equations can be written down for any given pure
state of a many-body system. By employing semi-classical radiation theory with these sources, the microscopic Maxwell equations can be used to compute the strong
radiation fields produced by interacting charged quantal particles. For a charged quantal particle, three radiation fields involve only the vector potential $\mathbf{A}$. This is another example demonstrating the observability of vector potential.
}{Five radiation fields are perpendicular to the canonical momentum of a single charged particle. For a group of charged particles, a new type of radiation field is predicted to be perpendicular to $\mathbf{A}(\mathbf{x}_{j},t)\times
\lbrack\nabla\times(\nabla_{j}\Psi^{\prime})]$, where $\Psi^{'}$ is the many-body wave function. This kind of radiation does not exist for a single charged particle. The macroscopic
Maxwell equations are derived from the corresponding microscopic equations for
a pure state by the Russakoff-Robinson procedure which requires only a
spatial coarse graining. Because the sources of fields are also the responses of a system to an external field, one also has to
give up the temporal coarse graining of the current density which is often supposed
to be critical in the kinetic approach of conductivity.
  }}

% The class file requires the standard graphicx Latex package. See the 'LaTeX
% standard graphics and color packages documentation' for more information at
% <http://tug.ctan.org/tex-archive/macros/latex/required/graphics/grfguide.pdf>.
%
% Accepted figure file formats depend on which LaTeX flavour is used.
% Classic LaTeX is always able to use Encapsulated Postscript (EPS);
% PDFLaTeX can't use this but accepts PDF, JPG, PNG, and GIF formats.
%
% See examples for implementing graphics in floating figure environments later in this file.
% If \titlefigure is given, it takes as its mandatory parameter the
% name (without extension) of some figure file.
%\titlefigure[height=3.1cm]{empty2w}
%\titlefigurecaption{%
%  This is the caption of the \emph{optional} abstract figure. If
%  there is no abstract figure here, the abstract text should be divided into both columns.
%}

\maketitle   % please do not remove

% Use the following code if you wish to generate your bibliography with BibTeX;
% replace the string "pss-demo" below with the name(s) of
% the BibTeX data base(s) you want to use.
% The resulting bibliography-output (the content of the .bbl file)
% must be pasted back into this file before submission.
% Please also include your BibTeX data base file(s) in your submission
% so that we can re-run BibTeX if necessary.
%
%\bibliographystyle{pss}
%\bibliography{pss-demo}
%
% Replace the following example bibliography with your references
% before submission:

\section{Introduction}

\label{intr}

This paper discusses three closely related problems for a group of
non-relativistic charged particles obeying quantum mechanics (QM): (1) the
radiation fields produced by the current density in a pure state of the
many-body system; (2) derivation of the macroscopic Maxwell equations from the
corresponding microscopic equations in a given pure state; and (3) the
consequence of the averaging procedure used in (2) to computing the electrical conductivity.

In condensed phases, there exist abundant radiation related phenomena.
Cherenkov radiation and transition radiation are two prominent examples\cite%
{lv8}. Although quantum electrodynamics (QED) is believed to be applicable
to all radiation problems, it is only feasible to calculate the weak field
(a few photons) produced by a few charged particles with perturbation theory%
\cite{hei,lv4,cla,dup,cra,sal}. QED is difficult for the long time evolution%
\cite{oud}, the reaction of the atom on the applied field\cite{cri,nni}, the
effect of boundary conditions\cite{akh}, bound states and coherent
radiation from many charged particles. To further explore radiation
phenomena in condensed phases, we need an alternative method for computing
field.

The ``neoclassical'' or semi-classical radiation theory (SCRT) of E. T. Jaynes\cite{jay} is
concerned with the strong radiation of a relativistic or non-relativistic
charged particle. The motion of the particle is treated quantum mechanically
while the electromagnetic field is treated classically\cite%
{oud,cri,nni,akh,jay}. The SCRT has succeeded in treating spontaneous
emission, absorption, induced emission\cite{nes,shi},\ photo-dissociation,
Raman scattering\cite{lee}, radiative level shifts\cite{bet} and absorption
of radiation by diffusive electron\cite{lig}. Thus it is natural to extend
SCRT to the radiation field caused by a current-carrying pure state of many
non-relativistic charged particles. In this paper, we merge results from the
microscopic response method (MRM)\cite{mic,long} with concepts of SCRT to
extend the latter to many-particle systems.

%For a macroscopic system, we do not know which pure state the system is initially in. Therefore we have to average over all possible initial states according to their statistical weights.
Since the pioneering work of Lorentz%
\cite{lor}, many schemes have been suggested to derive the macroscopic Maxwell
equations from the corresponding microscopic equations\cite{maz,ram,gro}. To
obtain the macroscopic fields and sources, usually both temporal coarse graining
and spatial coarse graining are taken in addition to ensemble average.
By assuming that the motion of charged particles obeys classical mechanics (CM),
Russakoff\cite{rus} and Robinson\cite{bins} (RR) proved that (i) only the
spatial coarse graining is relevant; (ii) the spatial coarse graining is
compatible with the ensemble average\cite{mab}. Since the motion of
conduction electrons in solids is described by QM, it is desirable to extend
Russakoff-Robinson's procedure to such a situation.

Dropping the temporal coarse graining has a profound consequence. In both
microscopic and macroscopic Maxwell equations, current density and charge
density are the sources of electromagnetic field. On the other hand, current
density is also the response of a system to an external field. Therefore
Russakoff-Robinson's procedure means that the temporal coarse-grained
average is irrelevant to derive the irreversibility caused by a conduction
process.

In Sec.\ref{max}, we write out the microscopic Maxwell equation for a pure
state. In Sec.\ref{tdao}, we apply SCRT to analyze the radiation fields
from the current density caused by one and many charged particles. For a quantal charged particle,
three radiation fields involve only the vector potential. Here we have one more example demonstrating the physical reality of vector potential. In contrast to the classical radiation field,
five radiation fields are perpendicular to the canonical momentum of particle. For a group of charged particles obeying quantum mechanics, one of its radiation fields is perpendicular to $\mathbf{A}(\mathbf{x}_{j},t)\times
\lbrack\nabla\times(\nabla_{j}\Psi^{\prime})]$, where $\Psi^{'}$ is the many-body wave function, $\mathbf{A} (\mathbf{x}_{j},t)$ is the vector potential felt by the $j^{th}$ particle at time t. This kind of radiation does not exist for a single charged particle. In Sec.\ref{derm}, we apply the RR\ procedure to derive
the macroscopic Maxwell equations for a group of charged particles obeying
QM. We show that without temporal coarse grained average, the
irreversibility of conduction process is still included in the present
theory.

\section{Microscopic Maxwell equations for a pure state}

\label{max}

Consider a system ($\mathcal{S}$) with $N_{e}$ electrons and $\mathcal{N}$
nuclei in an external electromagnetic field ($\mathcal{F}$) described by
vector and scalar potentials ($\mathbf{A},\phi$). Denote the coordinates of $%
N_{e}$ electrons as $\mathbf{r}_{1},\mathbf{r}_{2},\cdots,\mathbf{r}_{N_{e}}$%
, the coordinates of $\mathcal{N}$ nuclei as $\mathbf{W}_{1},\mathbf{W}%
_{2},\cdots,\mathbf{W}_{\mathcal{N}}$. In the external field ($\mathbf{A}%
,\phi$), the state $\Psi^{\prime}(t)$ of the system is determined by the
many-body Schr\"{o}dinger equation%
\begin{equation}
i\hbar\partial\Psi^{\prime}/\partial t=H^{\prime}\Psi^{\prime},   \label{ts}
\end{equation}
where $H^{\prime}(t)=H+H_{fm}(t)$ is the total Hamiltonian for $\mathcal{S}$+%
$\mathcal{F}$, $H$ is the Hamiltonian of system, $H_{fm}(t)$ is the
field-matter interaction:%
\begin{equation}
H_{fm}(t)=\sum_{j=1}^{N_{e}}\{\frac{i\hbar e}{2m}[\mathbf{A}(\mathbf{r}%
_{j},t)\cdot\nabla_{j}+\nabla_{j}\cdot\mathbf{A}(\mathbf{r}_{j},t)]
\label{fm}
\end{equation}%
\begin{equation*}
+\frac{e^{2}\mathbf{A}^{2}(\mathbf{r}_{j},t)}{2m}+e\phi(\mathbf{r}_{j},t)\}
\end{equation*}%
\begin{equation*}
+\sum_{L=1}^{\mathcal{N}}\{-\frac{i\hbar Z_{L}e}{2M_{L}}[\mathbf{A}(\mathbf{W%
}_{L},t)\cdot\nabla_{L}+\nabla_{L}\cdot\mathbf{A}(\mathbf{W}_{L},t)]
\end{equation*}%
\begin{equation*}
+\frac{(Z_{L}e)^{2}\mathbf{A}^{2}(\mathbf{W}_{L},t)}{2M_{L}}-Z_{L}e\phi(%
\mathbf{W}_{L},t)\}.
\end{equation*}
The time dependence of $H_{fm}(t)$ comes from the external field. The
arguments of $\Psi^{\prime}$ are $(\mathbf{r}_{1},\mathbf{r}_{2},\cdots,%
\mathbf{r}_{N_{e}}$; $\mathbf{W}_{1},\mathbf{W}_{2},\cdots,\mathbf{W}_{%
\mathcal{N}};t)$.

In QM, the state of the system at time $t$ is described by the solution $%
\Psi^{\prime}(t)$ of eq.(\ref{ts}), the microscopic charge density $%
\rho^{\Psi^{\prime}}$ and current density $\mathbf{j}^{\Psi^{\prime}}$ in
state $\Psi^{\prime}$ are completely determined by $\Psi^{\prime}(t)$. The
microscopic charge density $\rho^{\Psi^{\prime}}$ in state $\Psi^{\prime}$
is $\rho^{\Psi^{\prime}}(\mathbf{r},t)=\int d\tau\Psi^{\prime\ast}\widehat{%
\rho }(\mathbf{r})\Psi^{\prime}$, where $\widehat{\rho}(\mathbf{r}%
)=\sum_{j}e\delta(\mathbf{r}-\mathbf{r}_{j})-\sum_{L}Z_{L}e\delta(\mathbf{r}-%
\mathbf{W}_{L})$ is the charge density operator of the system, $d\tau =d%
\mathbf{r}_{1}d\tau^{1}$ is the volume element in the whole configurational
space, $d\tau^{1}=d\mathbf{r}_{2}\cdots d\mathbf{r}_{N_{e}}d\mathbf{W}%
_{1}\cdots d\mathbf{W}_{\mathcal{N}}$. Carrying out the integrals and using
the antisymmetry of $\Psi^{\prime}$ for exchanging the coordinates of two
electrons,
\begin{equation}
\rho^{\Psi^{\prime}}(\mathbf{r},t)=N_{e}e\int
d\tau^{1}\Psi^{\prime}\Psi^{\prime\ast}-\sum_{L}Z_{L}e\int
d\tau^{L}\Psi^{\prime}\Psi^{\prime\ast},   \label{md}
\end{equation}
where $d\tau^{L}=d\mathbf{r}_{1}\cdots d\mathbf{r}_{N_{e}}d\mathbf{W}%
_{1}\cdots d\mathbf{W}_{L-1}d\mathbf{W}_{L+1}\cdots d\mathbf{W}_{\mathcal{N}}
$, the arguments of the $\Psi^{\prime}$ in the first term are $(\mathbf{r},%
\mathbf{r}_{2}\cdots\mathbf{r}_{N_{e}}$; $\mathbf{W}_{1}\cdots\mathbf{W}_{%
\mathcal{N}})$, the arguments of the $\Psi^{\prime}$ in the second term are $%
(\mathbf{r}_{1},\cdots,\mathbf{r}_{N_{e}}$; $\mathbf{W}_{1},\cdots ,\mathbf{W%
}_{L-1},\mathbf{r},\mathbf{W}_{L+1}\cdots,\mathbf{W}_{\mathcal{N}};t)$.

\subsection{Microscopic Maxwell equations in a pure state}

If the system is in a\ state $\Psi ^{\prime }(t)$, the microscopic Maxwell
equations are%
\begin{equation}
\nabla \cdot \mathbf{b}^{\Psi ^{\prime }}=0,\text{ \ \ }\nabla \times
\mathbf{e}^{\Psi ^{\prime }}=-\partial \mathbf{b}^{\Psi ^{\prime }}/\partial
t,  \label{hom}
\end{equation}%
and

\begin{equation}
\nabla \cdot \mathbf{e}^{\Psi ^{\prime }}=\rho ^{\Psi ^{\prime }}/\epsilon
_{0}\text{, \ \ }c^{2}\nabla \times \mathbf{b}^{\Psi ^{\prime }}=\mathbf{j}%
^{\Psi ^{\prime }}/\epsilon _{0}+\partial \mathbf{e}^{\Psi ^{\prime
}}/\partial t,  \label{inh}
\end{equation}%
where $\mathbf{e}^{\Psi ^{\prime }}$ and $\mathbf{b}^{\Psi ^{\prime }}$ are
the microscopic electric field and magnetic induction at $(\mathbf{r},t)$ in
pure state $\Psi ^{\prime }$.

Applying $\partial/\partial t$ on the first inhomogeneous equation in (\ref%
{inh}), and $\nabla\cdot$ on the second, one has the continuity equation for
a pure state $\Psi^{\prime}$\cite{jac}:%
\begin{equation}
\partial\rho^{\Psi^{\prime}}(\mathbf{r},t)/\partial t+\nabla\cdot \mathbf{j}%
^{\Psi^{\prime}}(\mathbf{r},t)=0.   \label{chc}
\end{equation}
Eq.(\ref{chc}) helps us find the microscopic current density\cite{mic,long} $%
\mathbf{j}^{\Psi^{\prime}}$ from $\rho^{\Psi^{\prime}}$:%
\begin{equation*}
\mathbf{j}^{\Psi^{\prime}}(\mathbf{r},t)=\frac{i\hbar eN_{e}}{2m}\int
d\tau^{1}(\Psi^{\prime}\nabla\Psi^{\prime\ast}-\Psi^{\prime\ast}\nabla
\Psi^{\prime})
\end{equation*}%
\begin{equation}
-\frac{N_{e}e^{2}}{m}\mathbf{A}(\mathbf{r},t)\int d\tau^{1}\Psi^{\prime\ast
}\Psi^{\prime}   \label{urc}
\end{equation}%
\begin{equation*}
-\sum_{L=1}^{\mathcal{N}}\frac{i\hbar Z_{L}e}{2M_{L}}\int
d\tau^{L}(\Psi^{\prime}\nabla\Psi^{\prime\ast}-\Psi^{\prime\ast}\nabla\Psi^{%
\prime})
\end{equation*}%
\begin{equation*}
-\sum_{L=1}^{\mathcal{N}}\frac{(Z_{L}e)^{2}}{M_{L}}\mathbf{A}(\mathbf{r}%
,t)\int d\tau^{L}\Psi^{\prime\ast}\Psi^{\prime},
\end{equation*}
where the arguments of the $\Psi^{\prime}$ in the 1$^{st}$ and 2$^{nd}$
lines are $(\mathbf{r},\mathbf{r}_{2},\cdots,\mathbf{r}_{N_{e}}$; $\mathbf{W}%
_{1},\cdots,\mathbf{W}_{\mathcal{N}};t)$, the arguments of the $\Psi^{\prime}
$ in the 3$^{rd}$ and 4$^{th}$ lines is $(\mathbf{r}_{1},\cdots,\mathbf{r}%
_{N_{e}}$; $\mathbf{W}_{1},\cdots,\mathbf{W}_{L-1},\mathbf{r},\mathbf{W}%
_{L+1}\cdots,\mathbf{W}_{\mathcal{N}};t)$. In eq.(\ref{urc}),
the contribution to $\mathbf{j}^{\Psi^{\prime}}$ from nuclei is similar to
that from electrons; we will not keep the last two terms.

In the derivation of eq.(\ref{urc}) by either the principle of virtual work\cite{kubo} or the microscopic response method\cite{mic,long}, we deal with a specified external field in an arbitrary gauge. Therefore the current density (\ref{urc}) is valid for an arbitrary gauge. Because $(\phi,\mathbf{A})$  directly appear in Schrodinger equation (\ref{ts}), the many-body wave function depends on the chosen gauge. When we discuss radiation problems with coupled Maxwell equations (\ref{hom},\ref{inh}) and Schrodinger equation (\ref{ts}), it is natural to inquire whether fields, charge density and current density are affected in different gauges.
%For a group of charged particles obeying CM, the scalar and vector potentials are only two auxiliary quantities in solving Maxwell equations. However for a a group of charged particles obeying QM, the many-body wave function depends on the chosen gauge. It is important to reassure that the expressions for observable quantities such as fields, charge density and current density are correct in arbitrary gauge.
If one makes gauge transformation for potentials from ($\phi_{1}%
,\mathbf{A}_{1}$) to ($\phi_{2},\mathbf{A}_{2}$):%
\begin{equation}
\phi_{2}(\mathbf{r},t)=\phi_{1}(\mathbf{r},t)-\frac{\partial\chi
(\mathbf{r},t)}{\partial t},\label{g1}%
\end{equation}
and%
\begin{equation}
\mathbf{A}_{2}(\mathbf{r},t)=\mathbf{A}_{1}(\mathbf{r},t)+\nabla
\chi(\mathbf{r},t),\label{g2}%
\end{equation}
one has to replace
\begin{equation}
i\hbar\frac{\partial\Psi_{1}^{\prime}}{\partial t}=H^{\prime}[\phi
_{1},\mathbf{A}_{1}]\Psi_{1}^{\prime}\label{g3}%
\end{equation}
with
\begin{equation}
i\hbar\frac{\partial\Psi_{2}^{\prime}}{\partial t}=H^{\prime}[\phi
_{2},\mathbf{A}_{2}]\Psi_{2}^{\prime},\label{g4}%
\end{equation}
where%
\begin{equation}
\Psi_{2}^{\prime}(\mathbf{r}_{1},\mathbf{r}_{2},\cdots\mathbf{r}_{N}%
,t)=\Psi_{1}^{\prime}(\mathbf{r}_{1},\mathbf{r}_{2},\cdots\mathbf{r}%
_{N},t)\label{g5}%
\end{equation}
\[
\times\exp\{i\sum_{j=1}^{N}e_{j}\chi(\mathbf{r}_{j},t)/\hbar\}.
\]
Here $e_{j}$ is the charge of the $j^{th}$ charged particle.
Noticing the potentials felt by the $j^{th}$ particle are
\begin{equation}
\phi_{2}(\mathbf{r}_{j},t)=\phi_{1}(\mathbf{r}_{j},t)-\frac{\partial
\chi(\mathbf{r}_{j},t)}{\partial t}\label{g6}%
\end{equation}
and%
\begin{equation}
\mathbf{A}_{2}(\mathbf{r}_{j},t)=\mathbf{A}_{1}(\mathbf{r}_{j},t)+\nabla
_{\mathbf{r}_{j}}\chi(\mathbf{r}_{j},t),\label{g7}%
\end{equation}
one can easily reduce eq.(\ref{g4}) into eq.(\ref{g3}). With the hep of eqs.(\ref{g5},\ref{g6}\ref{g7}), we can directly prove that charge density (\ref{md}) and the current density calculated from eq.(\ref{urc}) are the same for any two different gauges. Since the sources of fields  $(\rho^{\Psi^{\prime}},\mathbf{j} ^{\Psi^{\prime}})$ in eqs.(\ref{hom},\ref{inh}) do not depend on gauge, the fields $(\mathbf{e} ^{\Psi^{\prime}},\mathbf{b} ^{\Psi^{\prime}})$ do not depend on gauge.

\subsection{Current density and dipole moment density}

The $\alpha^{th}$ ($\alpha=x,y,z$) component of the dipole density operator
is defined as%
\begin{equation}
\widehat{\rho}_{\alpha}^{d}(\mathbf{r})=\sum_{j=1}^{N_{e}}er_{j\alpha}\delta(%
\mathbf{r}-\mathbf{r}_{j})-\sum_{L=1}^{\mathcal{N}}Z_{L}eW_{L\alpha }\delta(%
\mathbf{r}-\mathbf{W}_{L}).   \label{dip}
\end{equation}
We may extract the polarization $P_{\alpha}^{\Psi^{\prime}}(\mathbf{r},t)$
in state $\Psi^{\prime}$ from%
\begin{equation}
d_{\alpha}(t)|_{\Psi^{\prime}}=\int d\mathbf{r}P_{\alpha}^{\Psi^{\prime}}(%
\mathbf{r},t),   \label{lar}
\end{equation}
where $d_{\alpha}(t)|_{\Psi^{\prime}}$ is the $\alpha^{th}$ component of the
induced dipole in state $\Psi^{\prime}$:%
\begin{equation}
d_{\alpha}(t)|_{\Psi^{\prime}}=\int d\mathbf{r}\int d\tau\Psi^{\prime\ast }%
\widehat{\rho}_{\alpha}^{d}\Psi^{\prime},\text{ }\alpha=x,y,z.   \label{avdi}
\end{equation}
The time dependence of the induced dipole in state $\Psi^{\prime}(t)$
results from the time dependence of $\Psi^{\prime}(t)$. The time derivative
of the polarization can be found from $\partial
d_{\alpha}(t)|_{\Psi^{\prime}}/\partial t$:%
\begin{equation}
\frac{\partial}{\partial t}d_{\alpha}(t)|_{\Psi^{\prime}}=\int d\mathbf{r}%
\frac{\partial}{\partial t}P_{\alpha}^{\Psi^{\prime}}(\mathbf{r},t).
\label{dao}
\end{equation}
Combining eqs.(\ref{ts},\ref{avdi},\ref{dao}) and integrating by parts, one
has:%
\begin{equation}
\frac{\partial P_{\alpha}^{\Psi^{\prime}}(\mathbf{r},t)}{\partial t}=\frac{%
N_{e}i\hbar e}{2m}\int d\tau^{1}(\Psi^{\prime}\frac{\partial
\Psi^{\prime\ast}}{\partial r_{\alpha}}-\Psi^{\prime\ast}\frac{\partial
\Psi^{\prime}}{\partial r_{\alpha}})   \label{pder}
\end{equation}%
\begin{equation*}
-\frac{N_{e}e^{2}}{m}A_{\alpha}(\mathbf{r},t)\int d\tau^{1}\Psi^{\prime\ast
}\Psi^{\prime}
\end{equation*}%
\begin{equation*}
-\sum_{L=1}^{\mathcal{N}}\frac{i\hbar Z_{L}e}{2M_{L}}\int
d\tau^{L}(\Psi^{\prime}\frac{\partial\Psi^{\prime\ast}}{\partial r_{\alpha}}%
-\Psi^{\prime\ast}\frac{\partial\Psi^{\prime}}{\partial r_{\alpha}})
\end{equation*}%
\begin{equation*}
-\sum_{L=1}^{\mathcal{N}}\frac{(Z_{L}e)^{2}}{M_{L}}A_{\alpha}(\mathbf{r}%
,t)\int d\tau^{L}\Psi^{\prime\ast}\Psi^{\prime},
\end{equation*}
where the arguments of $\Psi^{\prime}$ in the 1$^{st}$ and 2$^{nd}$
lines are $(\mathbf{r},\mathbf{r}_{2},\cdots,\mathbf{r}_{N_{e}};\mathbf{W}%
_{1},\cdots,\mathbf{W}_{\mathcal{N}};t)$, the arguments of the $\Psi^{\prime}
$ in the 3$^{rd}$ and 4$^{th}$ lines is $(\mathbf{r}_{1},\cdots,\mathbf{r}%
_{N_{e}}$; $\mathbf{W}_{1},\cdots,\mathbf{W}_{L-1},\mathbf{r},\mathbf{W}%
_{L+1}\cdots,\mathbf{W}_{\mathcal{N}};t)$. Comparing eq.(\ref{pder}) with
eq.(\ref{urc}), we find that the current density in state $\Psi^{\prime}$ is
related to the time derivative of polarization in state $\Psi^{\prime}$ by:%
\begin{equation}
j_{\alpha}^{\Psi^{\prime}}(\mathbf{r},t)=\frac{\partial
P_{\alpha}^{\Psi^{\prime}}(\mathbf{r},t)}{\partial t},\text{ }\alpha=x,y,z.
\label{cur}
\end{equation}
Since both the free carriers and the bound electrons are included in eq.(\ref%
{dip}), the polarization $P_{\alpha}^{\Psi^{\prime}}$ defined in eq.(\ref%
{lar}) contains the contributions from both carriers and bound electrons.

\subsection{Radiation field}

The state of $\mathcal{S}$+$\mathcal{F}$ is determined by the
coupled equations (\ref{ts},\ref{hom},\ref{inh})\cite{shi}. In ordinary materials, the
induced motion of the charged particles by a weak external field is
non-relativistic: the energy radiated by a particle is much smaller than the
mechanical energy of that particle. Using direct product $|\Psi^{\prime
}\rangle\otimes|\mathbf{e},\mathbf{b}\rangle$ to represent the state of $%
\mathcal{S}$+$\mathcal{F}$ is then allowed.

Although the vector potential appears in the expression (\ref{urc}) of current density, but eq.(\ref{urc}) is correct to arbitrary gauge\cite{mic,long}. In the same spirit, we present a method calculating field which does not refer potentials.
To obtain the wave equation for $\mathbf{b}^{\Psi^{\prime}}$, one applies
$\nabla\times$ to the Ampere-Maxwell law, and notices the magnetic Gauss
theorem and the Faraday-Lenz's laws. One has\cite{shi}%
\begin{equation}
\nabla^{2}\mathbf{b}^{\Psi^{\prime}}-\frac{1}{c^{2}}\frac{\partial
^{2}\mathbf{b}^{\Psi^{\prime}}}{\partial t^{2}}=-\mu_{0}\nabla\times
\mathbf{j}^{\Psi^{\prime}}.\label{wq1}%
\end{equation}
The wave equation for $\mathbf{e}^{\Psi^{\prime}}$ can be obtained by applying
$\nabla\times$ to the the Faraday-Lenz's laws, and noticing the electric Gauss
theorem and the the Ampere-Maxwell law. One has
\begin{equation}
\nabla^{2}\mathbf{e}^{\Psi^{\prime}}-\frac{1}{c^{2}}\frac{\partial
^{2}\mathbf{e}^{\Psi^{\prime}}}{\partial t^{2}}=\frac{\nabla\rho^{\Psi
^{\prime}}}{\epsilon_{0}}+\mu_{0}\frac{\partial\mathbf{j}^{\Psi^{\prime}}%
}{\partial t}.\label{wq2}%
\end{equation}
The temporal Fourier transformations of eqs.(\ref{chc},\ref{wq1},\ref{wq2}) are%
\begin{equation}
-i\omega\rho_{\omega}^{\Psi^{\prime}}+\mathbf{j}_{\omega}^{\Psi^{\prime}%
}=0,\label{ct1}%
\end{equation}
and%
\begin{equation}
\nabla^{2}\mathbf{b}_{\omega}^{\Psi^{\prime}}+k^{2}\mathbf{b}_{\omega}%
^{\Psi^{\prime}}=-\mu_{0}\nabla\times\mathbf{j}_{\omega}^{\Psi^{\prime}%
},\text{ \ }k=\omega/c,\label{wq10}%
\end{equation}
and%
\begin{equation}
\nabla^{2}\mathbf{e}_{\omega}^{\Psi^{\prime}}+k^{2}\mathbf{e}_{\omega}%
^{\Psi^{\prime}}=\frac{\nabla\rho_{\omega}^{\Psi^{\prime}}}{\epsilon_{0}%
}-i\omega\mu_{0}\mathbf{j}_{\omega}^{\Psi^{\prime}}.\label{wq20}%
\end{equation}
By means of the Green's function method\cite{pan,jac}, the retarded solutions
of eqs.(\ref{wq10},\ref{wq20}) are
\begin{equation}
\mathbf{b}_{\omega}^{\Psi^{\prime}}(\mathbf{x})=\frac{\mu_{0}}{4\pi}\int
d^{3}x^{\prime}\frac{e^{ikr}}{r}\nabla_{\mathbf{x}^{\prime}}\times
\mathbf{j}_{\omega}^{\Psi^{\prime}}(\mathbf{x}^{\prime}),\label{wq11}%
\end{equation}
and%
\begin{equation}
\mathbf{e}_{\omega}^{\Psi^{\prime}}=\frac{1}{4\pi\epsilon_{0}}\int
d^{3}x^{\prime}\frac{e^{ikr}}{r}\{\frac{i\omega}{c^{2}}\mathbf{j}_{\omega
}^{\Psi^{\prime}}(\mathbf{x}^{\prime})-\nabla_{\mathbf{x}^{\prime}}%
\rho_{\omega}^{\Psi^{\prime}}(\mathbf{x}^{\prime})\},\label{wq21}%
\end{equation}
where $\mathbf{r}=\mathbf{x}-\mathbf{x}^{\prime}$ and\ $r=|\mathbf{r}|$.
Integrating by parts and noticing eq.(\ref{ct1}), one has%
\begin{equation}
\mathbf{b}_{\omega}^{\Psi^{\prime}}(\mathbf{x})=\frac{\mu_{0}}{4\pi}\int
d^{3}x^{\prime}\frac{e^{ikr}}{r^{2}}\{\frac{\mathbf{j}_{\omega}^{\Psi^{\prime
}}\times\mathbf{r}}{r}-ik\mathbf{j}_{\omega}^{\Psi^{\prime}}\times
\mathbf{r}\},\label{wq12}%
\end{equation}
and%
\begin{equation}
4\pi\epsilon_{0}\mathbf{e}_{\omega}^{\Psi^{\prime}}(\mathbf{x})=\frac{ik}%
{c}\int d^{3}x^{\prime}\frac{e^{ikr}}{r^{3}}[\mathbf{r}\times(\mathbf{j}%
_{\omega}^{\Psi^{\prime}}\times\mathbf{r})]\label{wq22}%
\end{equation}%
\[
+\int d^{3}x^{\prime}\frac{e^{ikr}}{r^{3}}\{\mathbf{r}\rho_{\omega}%
^{\Psi^{\prime}}(\mathbf{x}^{\prime})+\frac{1}{cr}[(\mathbf{j}_{\omega}%
^{\Psi^{\prime}}\cdot\mathbf{r})\mathbf{r}-\mathbf{r}\times(\mathbf{j}%
_{\omega}^{\Psi^{\prime}}\times\mathbf{r})]\}.
\]
To obtain eqs.(\ref{wq12},\ref{wq22}), traditionally one first introduces scalar and vector potentials. The wave equations for potentials are then derived under a given gauge, e.g. Lorentz gauge. One arrives fields by differentiating the potentials. Of course, eqs.(\ref{era},\ref{scf}) do not depend on any choice of gauge of the potentials\cite{pan}.

Taking inverse Fourier transformation, the radiation fields produced by a current-carrying pure state $\Psi^{\prime
}(t)$ are\cite{pan}:%
\begin{equation}
\mathbf{b}_{rad}^{\Psi^{\prime}}(\mathbf{x},t)=\frac{1}{4\pi\epsilon_{0}c^{3}%
}\int d^{3}x^{\prime}\frac{[\frac{\partial\mathbf{j}^{\Psi^{\prime}}}{%
\partial t^{\prime}}(\mathbf{x}^{\prime},t^{\prime})]_{ret}\times(\mathbf{x}-%
\mathbf{x}^{\prime})}{|\mathbf{x}-\mathbf{x}^{\prime}|^{2}},   \label{era}
\end{equation}

\begin{equation}
\mathbf{e}_{rad}^{\Psi^{\prime}}(\mathbf{x},t)=\frac{1}{4\pi\epsilon_{0}c^{2}%
}\int d^{3}x^{\prime}   \label{scf}
\end{equation}%
\begin{equation*}
\frac{\{[\frac{\partial\mathbf{j}^{\Psi^{\prime}}}{\partial t^{\prime}}(%
\mathbf{x}^{\prime},t^{\prime})]_{ret}\times(\mathbf{x}-\mathbf{x}^{\prime
})\}\times(\mathbf{x}-\mathbf{x}^{\prime})}{|\mathbf{x}-\mathbf{x}^{\prime
}|^{3}},
\end{equation*}
where $t^{\prime}=t-|\mathbf{x}-\mathbf{x}^{\prime}|/c$. It
is the time derivative $\partial\mathbf{j}^{\Psi^{\prime}}(\mathbf{x}%
^{\prime },t^{\prime})/\partial t^{\prime}$ of the current density that
determines the radiation field. Although $\nabla\times\mathbf{j}_{\omega}^{\Psi^{\prime}}$ and $\nabla\rho^{\Psi
^{\prime}}$ also appear as the sources of fields in eqs.(\ref{wq1},\ref{wq2}), making use of the integration by parts and eq.(\ref{ct1}), they are partly converted into $\partial{\mathbf{j}^{\Psi^{\prime}}}/\partial{t^{\prime}}$ (contributed to radiation field) and partly converted into $\mathbf{j}^{\Psi^{\prime}}$ and $\rho$ (contributed to the induction field), cf. eqs.(\ref{wq11},\ref{wq21}).

Denote $\mathbf{x}_{0}^{\prime }$ as the origin of the source distribution, $%
\mathbf{x}^{\prime }$ an arbitrary point in a localized charge distribution,
$\mathbf{x}$ the observation point. The integrands of eqs.(\ref{era},\ref%
{scf}) can be expanded in a small parameter $\varepsilon =|\mathbf{x}^{\prime
}-\mathbf{x}_{0}^{\prime }|/|\mathbf{x}-\mathbf{x}_{0}^{\prime }|$. Because
the radiation field represents outgoing energy, it must have asymptotic form  proportional to $|\mathbf{x}-\mathbf{x}_{0}^{\prime }|^{-1}$\cite{jac,pan}. The electric
field and the magnetic induction in far region are%
\begin{equation}
\mathbf{b}_{rad}^{\Psi ^{\prime }}=-\mathbf{n}\times \mathbf{p}^{\Psi
^{\prime }},\text{ \ \ }\mathbf{e}_{rad}^{\Psi ^{\prime }}=c\mathbf{n}\times
(\mathbf{n}\times \mathbf{p}^{\Psi ^{\prime }}),  \label{eb}
\end{equation}%
where $\mathbf{n}=(\mathbf{x}-\mathbf{x}_{0}^{\prime })/|\mathbf{x}-\mathbf{x%
}_{0}^{\prime }|$, the arguments of $\mathbf{e}_{rad}^{\Psi ^{\prime }}$, $%
\mathbf{b}_{rad}^{\Psi ^{\prime }}$ and $\mathbf{p}^{\Psi ^{\prime }}$ are $(%
\mathbf{x},t)$\cite{v2,dav}. The vector field $\mathbf{p}^{\Psi ^{\prime }}$ is given by
\begin{equation}
\mathbf{p}^{\Psi ^{\prime }}(\mathbf{x},t)=\frac{1}{4\pi \epsilon _{0}c^{3}}%
\int d^{3}x^{\prime }  \label{mp}
\end{equation}%
\begin{equation*}
\sum_{q=0}^{\infty }\frac{(\frac{|\mathbf{x}^{\prime }-\mathbf{x}%
_{0}^{\prime }|}{c}\cos \theta )^{q}}{q!}\frac{\partial ^{q+1}\mathbf{j}%
^{\Psi ^{\prime }}(\mathbf{x}^{\prime },t-s_{0})}{\partial t^{q+1}},
\end{equation*}%
where $\theta $ is the angle between $(\mathbf{x}-\mathbf{x}_{0}^{\prime })$
and $(\mathbf{x}^{\prime }-\mathbf{x}_{0}^{\prime })$. The $q=0$ term gives
the dipole approximation, the $q=1$ term gives the quadrupole and magnetic
dipole approximation etc\cite{v2,dav}.

\section{Radiation fields computed from
$\partial\mathbf{j}^{\Psi^{\prime}}(\mathbf{x}^{\prime},t^{\prime})/\partial
t^{\prime}$}

\label{tdao}

\subsection{A moving electron}

\label{tdao1}

Let us consider the motion of an electron in an electromagnetic field ($%
\mathbf{A},\phi$) and another external field $U^{\prime}(\mathbf{x})$. An
alkali atom in a dilute alkali gas is an example. Denote the single particle wave function of
the particle as $\psi^{\prime}(\mathbf{x},t)$, the average velocity $%
\overline{\mathbf{v}}(t)|_{\psi^{\prime}}=d[\int d^{3}x\psi^{\prime\ast }%
\mathbf{r}\psi^{\prime}]/dt$ in state $\psi^{\prime}(\mathbf{x},t)$ can be
found from Schrodinger equation\cite{shi,mic}:%
\begin{equation*}
\overline{\mathbf{v}}(t)|_{\psi^{\prime}}=\frac{i\hbar}{2m}\int
d^{3}x(\psi^{\prime}\nabla\psi^{\prime\ast}-\psi^{\prime\ast}\nabla\psi^{%
\prime})
\end{equation*}%
\begin{equation}
-\frac{e}{m}\int d^{3}x\psi^{\prime\ast}\mathbf{A}(\mathbf{x},t)\psi^{\prime
}.   \label{vv}
\end{equation}
The average acceleration $\overline{\mathbf{a}}(t)|_{\psi^{\prime}}=d%
\overline{\mathbf{v}}(t)|_{\psi^{\prime}}/dt$ in state $\psi^{\prime }(%
\mathbf{x},t)$ is\cite{shi,mic}%
\begin{equation*}
\overline{\mathbf{a}}(t)|_{\psi^{\prime}}=m^{-1}\int d^{3}x\psi^{\prime\ast
}[e\mathbf{E}-\nabla U^{\prime}]\psi^{\prime}
\end{equation*}%
\begin{equation*}
+\frac{e}{m^{2}}\int d^{3}x\psi^{\prime\ast}(-i\hbar\nabla-e\mathbf{A}%
)\psi^{\prime}\times\mathbf{B}
\end{equation*}%
\begin{equation}
-\frac{i\hbar e}{2m^{2}}\int d^{3}x\psi^{\prime\ast}\psi^{\prime}(\nabla
\times\mathbf{B}),   \label{aa}
\end{equation}
where the arguments of the fields are $(\mathbf{x},t)$. The first term in eq.(%
\ref{aa}) is the acceleration caused by the electric field and external field $%
U^{\prime}$, the second term is the magnetic force. They are expected
from CM. The third term is a quantum effect: an additional small component
along the average direction of $\nabla\times\mathbf{B}$. The ratio of the
third term to the second is $\sim a/L$, where $a$ is the characteristic
length of wave function $\psi^{\prime}$, $L$ is the characteristic length scale in which
magnetic field changes.

For a single electron, eq.(\ref{urc}) reduces to:
\begin{equation*}
\mathbf{j}^{\psi^{\prime}}(\mathbf{x},t)=\frac{i\hbar e}{2m}(\psi^{\prime
}\nabla\psi^{\prime\ast}-\psi^{\prime\ast}\nabla\psi^{\prime})
\end{equation*}%
\begin{equation}
-\frac{e^{2}}{m}\mathbf{A}(\mathbf{x},t)\psi^{\prime\ast}\psi^{\prime}.
\label{1v}
\end{equation}
Comparing eq.(\ref{vv}) and eq.(\ref{1v}), we have:
\begin{equation}
\overline{\mathbf{v}}(t)|_{\psi^{\prime}}=e^{-1}\int d^{3}x\mathbf{j}%
^{\psi^{\prime}}(\mathbf{x},t).   \label{vc}
\end{equation}
The average acceleration and the time derivative $\partial\mathbf{j}%
^{\psi^{\prime}}(\mathbf{x},t)/\partial t$ of current density are related
by:\
\begin{equation}
\overline{\mathbf{a}}(t)|_{\psi^{\prime}}=e^{-1}\int d^{3}x\frac {\partial%
\mathbf{j}^{\psi^{\prime}}(\mathbf{x},t)}{\partial t}.   \label{adc}
\end{equation}

From eqs.(\ref{mp},\ref{adc}) we can see that in the \textit{dipole approximation} the quantal acceleration $\overline {%
\mathbf{a}}(t)|_{\psi^{\prime}}$ is the source of radiation. The radiation fields produced by the first two terms in eq.(\ref{aa}) are already known in CM. The last term in
eq.(\ref{aa}) is the new feature of a quantal charged particle. The corresponding radiation fields are perpendicular to $(\nabla
\times\mathbf{B})$, they should be detectable for a charged particle moving in a non-uniform magnetic
field.
By testing the polarization of scattered fields, it can also be observed in the scattering phenomenon.

Let us consider more general radiation fields produced by a quantal charged particle moving in an external field ($\phi,\mathbf{A}$). Since we consider only non-relativistic motion, the effect of self-fields of the particle on its motion can be neglected.
Taking time derivative in eq.(\ref{1v}), one finds
\begin{equation*}
\partial\mathbf{j}^{\psi^{\prime}}(\mathbf{x},t)/\partial t=\frac{e^{2}}{m}%
\mathbf{E}\psi^{'}\psi^{'*}-\frac{e}{m}\nabla U^{\prime}(\mathbf{x})\psi^{\prime}\psi^{\prime\ast}
\end{equation*}%
\begin{equation*}
+\frac{e^{2}}{2m^{2}}[\psi^{\prime}(i\hbar\nabla\psi^{\prime\ast})+\psi^{%
\prime\ast}(-i\hbar\nabla\psi^{\prime})]\times\mathbf{B}-\frac{e^{3}}{m^{2}}(%
\mathbf{A}\times\mathbf{B})\psi^{\prime}\psi^{\prime\ast}
\end{equation*}%
\begin{equation*}
+\frac{e}{m}\mathbf{A}(\nabla\cdot\mathbf{j}^{\psi^{\prime}})-\frac{e^{3}}{%
m^{2}}\psi^{\prime}\psi^{\prime\ast}(\mathbf{A}\cdot\nabla)\mathbf{A}
\end{equation*}%
\begin{equation*}
+\frac{i\hbar e^{2}}{2m^{2}}\{\psi^{\prime}[(\nabla\psi^{\prime\ast})\cdot%
\nabla]\mathbf{A}-\psi^{\prime\ast}[(\nabla\psi^{\prime})\cdot \nabla]%
\mathbf{A}\}
\end{equation*}%
\begin{equation*}
+\frac{\hbar^{2}e}{4m^{2}}[\psi^{\prime}\nabla\nabla^{2}\psi^{\prime\ast}+%
\psi^{\prime\ast}\nabla\nabla^{2}\psi^{\prime}-(\nabla\psi^{\prime\ast
})\nabla^{2}\psi^{\prime}-(\nabla\psi^{\prime})\nabla^{2}\psi^{\prime\ast}]
\end{equation*}%
\begin{equation*}
+\frac{i\hbar e^{2}}{2m^{2}}\{\psi^{\prime}[\mathbf{A}\cdot\nabla](\nabla
\psi^{\prime\ast})-\psi^{\prime\ast}[\mathbf{A}\cdot\nabla](\nabla\psi
^{\prime})]\}
\end{equation*}%
\begin{equation*}
+\frac{i\hbar e^{2}}{2m^{2}}[(\nabla\psi^{\prime\ast})\mathbf{A}\cdot
\nabla\psi^{\prime}-(\nabla\psi^{\prime})\mathbf{A}\cdot\nabla\psi^{\prime
\ast}]
\end{equation*}%
\begin{equation}
+\frac{i\hbar e^{2}}{2m^{2}}(\psi^{\prime}\nabla\psi^{\prime\ast}-\psi
^{\prime\ast}\nabla\psi^{\prime})\nabla\cdot\mathbf{A},   \label{1ct}
\end{equation}
where $\mathbf{E}(\mathbf{x},t)=-\nabla\phi(\mathbf{x},t)-\partial \mathbf{A}%
(\mathbf{x},t)/\partial t$ and $\mathbf{B}(\mathbf{x},t)=\nabla \times%
\mathbf{A}(\mathbf{x},t)$ are the electric field and magnetic induction, the
arguments of $\mathbf{A}$, $\mathbf{B}$, $\mathbf{E}$ and $\psi^{\prime}$
are $(\mathbf{x},t)$. Integrating eq.(\ref{1ct}) over space $\int d^{3}x$,
one can check sum rule (\ref{adc}) using integration by parts. However one
observes that, according to eqs.(\ref{era},\ref{scf}), the radiation
field is determined by the time derivative of $\partial\mathbf{j}%
^{\psi^{\prime}}(\mathbf{x},t)/\partial t$ current density rather than by
the average acceleration $\overline{\mathbf{a}}(t)$.

The radiation fields produced by the first three terms are perpendicular to
the electric force, external force of $U^{\prime}$, and magnetic force
respectively. They are well known in the classical description\cite{jac,pan,v2}.

The remaining terms of eq.(\ref{1ct}) result from the quantum features of the microscopic current density $\mathbf{j}^{\psi^{\prime}}$ in a pure state $\psi^{\prime}$. According to
eqs.(\ref{era},\ref{scf}), the radiation fields produced by the fourth term
are perpendicular to $(\mathbf{A}\times\mathbf{B})$, the radiation fields
produced by the fifth term are perpendicular to $\mathbf{A}$.
The radiation fields produced by the sixth terms are perpendicular to
$(\mathbf{A}\cdot\nabla)\mathbf{A}$. The ratio of the fourth term to the
second term is $\lambda/a$, where $\lambda$ is the characteristic length of
$\mathbf{A}$. The fifth term is the same order of magnitude as the third term.
The ratio of the fifth term to the second term is $A/(\hbar/ea)$, for ordinary
magnetic field this is a small number. The sixth term is the same order of
magnitude as the fourth term.

The radiation fields produced by the seventh term are perpendicular to
$[(\nabla\psi^{\prime})\cdot\nabla]\mathbf{A}$. The ratio of the seventh term
to the second term is $\sim(\hbar/ea^{2})/B$. The radiation fields
produced by the eighth term are perpendicular to the canonical momentum $\nabla\psi^{\prime
}$. The ratio of the eighth term to the second term is $\sim$1. In CM, no
such a radiation field exist. The radiation field produced by the ninth terms
are perpendicular to $[\mathbf{A}\cdot\nabla](\nabla\psi^{\prime})$. The ratio
of the eighth and the ninth terms to the second term is $\lambda/a$.

The radiation fields produced by the tenth term of eq.(\ref{1ct}) are also
perpendicular to $(\nabla\psi^{\prime})\mathbf{A}\cdot\nabla\psi^{'*}$, where $-i\hbar\nabla=m\mathbf{v}+e\mathbf{A}$ is the canonical momentum operator. In contrast to the eighth
term, which does not depend on vector potential, the tenth term depends on
$\mathbf{A}\cdot\nabla\psi^{\prime}$. The ratio of the tenth term to the fifth
term is $\sim$1. Both the fifth and the tenth terms exist for a constant
vector potential. The radiation fields produced by the eleventh term are perpendicular to $(\psi^{'}\nabla\psi^{'*})\nabla\cdot\mathbf{A}$.
The eleventh term is the same order of magnitude as the seventh term.

Before the Bohm-Aharonov effect was discovered, it had been assumed that any ($\phi^{'},\mathbf{A}^{'}$) is indistinguishable from ($\phi,\mathbf{A}$) if they satisfy $\mathbf{A}^{'}=\mathbf{A}+\nabla\Lambda$ and $\phi^{'}=\phi+\partial \Lambda / \partial t$, where is $\Lambda$ is an arbitrary scalar function of time and coordinate. The Bohm-Aharonov effect gives the first example that beside $\mathbf{B}=\nabla\times\mathbf{A}$, vector potential can be detected in another way: the phase shift $\frac{e}{\hbar}\oint\mathbf{A}\cdot d\mathbf{x}$. The radiation fields produced by the fourth, fifth, sixth, ninth, tenth and eleventh terms involve either $\mathbf{A}$ or $\nabla\cdot\mathbf{A}$. They demonstrate the observability of vector potential: one may detect these radiation fields outside a long solenoid
for which $\mathbf{B}$=0 but $\mathbf{A}\neq$0.

Eq.(\ref{1ct}) is a resolution of $\partial\mathbf{j}^{\Psi^{\prime}}(\mathbf{x}^{\prime},t^{\prime})/\partial t^{\prime}$. It directly results from the time dependent Schrodinger equation and eq.(\ref{1v}).
One may wonder whether a regrouping the terms in eq.(\ref{1ct}) leads to a different explanation for the radiation fields. The answer is no: because eqs.(\ref{era},\ref{scf}) are linear in $\partial\mathbf{j}^{\Psi^{\prime}}(\mathbf{x}^{\prime},t^{\prime})/\partial t^{\prime}$, a recombination of terms only results to a different choice of the basis of fields. The new basis of field is a superposition of the old fields.

In a pure state $\psi^{\prime}$, the fields produced by the 11 terms in eq.(\ref{1ct}) are coherent: during the whole motion of particle, all the 11 field components are in phase. But for a mixed state (e.g. vapor of Na atoms in the same excited state), the various terms in eq.(\ref{1ct}) are no longer coherent for a given atom. However for the long wave scattering, the same type term for neighboring atoms is coherent. This is similar to the scattering produced by a group of classical charged particles.
\subsection{A group of charged particles}

\label{mta}

Taking the time derivative in eq.(\ref{urc}), we obtain
\begin{equation}
\partial\mathbf{j}^{\Psi^{\prime}}(\mathbf{r},t)/\partial t=\frac{e^{2}}%
{m}\mathbf{E}(\mathbf{r},t)n^{\prime}(\mathbf{r},t)\label{ace}%
\end{equation}%
\[
+\frac{eN_{e}}{m}\int d\tau^{1}\Psi^{\prime}\Psi^{\prime\ast}[-\sum_{k}\nabla
U(\mathbf{r},\mathbf{x}_{k})-\sum_{\alpha}\nabla U^{\prime}(\mathbf{r}%
,\mathbf{W}_{\alpha})]
\]%
\[
+\frac{e^{2}N_{e}}{2m^{2}}\{\int d\tau^{1}[\Psi^{\prime}(i\hbar\nabla
\Psi^{\prime\ast})+\Psi^{\prime\ast}(-i\hbar\nabla\Psi^{\prime})]\}\times
\mathbf{B}(\mathbf{r},t)
\]%
\[
-\frac{e^{3}}{m^{2}}[\mathbf{A}\times\mathbf{B]}n^{\prime}+\frac{e}%
{m}\mathbf{A}(\mathbf{r},t)(\nabla\cdot\mathbf{j}^{\Psi^{\prime}})-\frac
{e^{3}}{m^{2}}[(\mathbf{A}\cdot\nabla)\mathbf{A]}n^{\prime}%
\]%
\[
+\frac{i\hbar e^{2}N_{e}}{2m^{2}}\int d\tau^{1}\{\Psi^{\prime}[(\nabla
\Psi^{\prime\ast})\cdot\nabla]\mathbf{A}(\mathbf{r},t)
\]%
\[
-\Psi^{\prime\ast}[(\nabla\Psi^{\prime})\cdot\nabla]\mathbf{A}(\mathbf{r},t)\}
\]%
\[
+\frac{\hbar^{2}eN_{e}}{4m^{2}}\int d\tau^{1}[\Psi^{\prime}\nabla\nabla
^{2}\Psi^{\prime\ast}+\Psi^{\prime\ast}\nabla\nabla^{2}\Psi^{\prime}%
\]%
\[
-(\nabla\Psi^{\prime\ast})\nabla^{2}\Psi^{\prime}-(\nabla\Psi^{\prime}%
)\nabla^{2}\Psi^{\prime\ast}]
\]%
\[
+\frac{i\hbar e^{2}N_{e}}{2m^{2}}\sum_{j}\int d\tau^{1}\{\Psi^{\prime
}[\mathbf{A}(\mathbf{x}_{j},t)\cdot\nabla](\nabla_{j}\Psi^{\prime\ast})
\]%
\[
-\Psi^{\prime\ast}[\mathbf{A}(\mathbf{x}_{j},t)\cdot\nabla](\nabla_{j}%
\Psi^{\prime})]\}
\]%
\[
+\frac{i\hbar e^{2}N_{e}}{2m^{2}}\sum_{j}\int d\tau^{1}[(\nabla\Psi
^{\prime\ast})\mathbf{A}(\mathbf{x}_{j},t)\cdot\nabla_{j}\Psi^{\prime}%
\]%
\[
-(\nabla\Psi^{\prime})\mathbf{A}(\mathbf{x}_{j},t)\cdot\nabla_{j}\Psi
^{\prime\ast}]
\]%
\[
+\frac{i\hbar N_{e}e^{2}}{2m^{2}}\int d\tau^{1}(\Psi^{\prime}\nabla
\Psi^{\prime\ast}-\Psi^{\prime\ast}\nabla\Psi^{\prime})\sum_{j}\nabla_{j}%
\cdot\mathbf{A}(\mathbf{x}_{j},t)
\]%
\[
+\frac{i\hbar e^{2}N_{e}}{2m^{2}}\int d\tau^{1}\sum_{j}\mathbf{A}%
(\mathbf{x}_{j},t)\times\{[\nabla\times(\nabla_{j}\Psi^{\prime\ast}%
)]\Psi^{\prime}%
\]%
\[
-[\nabla\times(\nabla_{j}\Psi^{\prime})]\Psi^{\prime\ast}\},
\]
where $\nabla_{\mathbf{r}}$ is abbreviated as $\nabla$. $n^{\prime}(\mathbf{r%
},t)=N_{e}\int d\tau\prime\Psi^{\prime}\Psi^{\prime\ast}$ is the number
density of electrons in state $\Psi^{\prime}$. $U$ and $U^{\prime}$ are the
interaction between two electrons and the interaction between an electron
and a nucleus. Because the contributions from nuclei are less important and
do not have any new features for visible light, they are not included in eq.(\ref{ace}). Of course the nuclear contributions are important for infrared radiation.

There are 12 terms in eq.(\ref{ace}). The first to
eleventh terms are a many-body generalization of the corresponding
single-particle terms in eq.(\ref{1ct}). The coherent scattering in CM is produced by
several particles in which their positions are correlated in the range of one
wavelenth\cite{jac,pan}. This feature is also inherited by the first to
eleventh terms. We don't have sum rules for the system with more than one
charged particles.

The twelfth term of eq.(\ref{ace}) exits only for a system with more than one
charged particle which obey QM, and represents a new feature of the radiation field
produced by the motion of many charged particles. If there is only one
particle, $\nabla\times(\nabla_{j}\Psi^{\prime})=0$, eq.(\ref{ace}) is
reduced to eq.(\ref{1ct}). The radiation fields emitted by the twelfth term
of eq.(\ref{ace}) are perpendicular to $\mathbf{A}(\mathbf{x}_{j},t)\times
\lbrack\nabla\times(\nabla_{j}\Psi^{\prime})]$, and are different to those
emitted by the first to the eleventh terms. The ratio of the twelfth term to
the magnetic force is $\lambda/a^{\prime}$, where $a^{\prime}$ is the characteristic
length of $\Psi^{\prime}$. It should be detectable for the scattering of
visible light by particles smaller than 100nm.

In QED, the coupling between electromagnetic field and charged particles is treated as a perturbation. The observable photons are expressed by external lines. To calculate a process with $n$ photons, one needs a $n^{th}$ order transition amplitude\cite{hei,lv4}. This is impractical for a strong field (many photons) produced by a large number of interacting charged particles. In SCRT, $H_{fm}$ is treated at zero order. Because no high energy photon appears in the usual state of a condensed phase and nanoscale system, eqs.(\ref{era},\ref{scf},\ref{ace}) are suitable for the emission and  scattering of light\cite{jay,nes,shi,lee,bet}.

If a system has only a few charged particles (or only a few charged particle play important role), the coherence of radiation is the same as the situation for a single charged particle: i.e. for a pure state, all the terms are coherent; For a mixed state, various terms in eq.(\ref{ace}) are no longer coherent. But for long wave scattering, the neighboring molecules are coherent for the same type of term in eq.(\ref{ace}).  If a large number of charged particles contribute to radiation, the macroscopic many-body pure state itself cannot exist for a long period\cite{vis}. The coherence between the terms in eq.(\ref{ace}) cannot be maintained in a macroscopic time scale (above micron second). For each term in eq.(\ref{ace}), one may have super-radiation phenomenon\cite{dick}.
\section{Macroscopic Maxwell equations}

\label{derm}

For a nanoscale system or a macroscopic system, we usually
 cannot specify the initial condition of the system precisely in the sense
that the state of the system in the future can be predicted to the maximum extent
allowed by QM\cite{tol}.
The results obtained in Sec.\ref{tdao} are not directly applicable. To explore radiation fields produced by a macroscopic or nanoscale system, we have to average the above results over a representative ensemble. In other words, we need the macroscopic Maxwell equations for a group of charged particles obeying QM.

Suppose that the system is in a good thermal contact with a reservoir ($\mathcal{B}$)
such that after a short equilibration  time, the system reaches a steady state:
 the thermodynamic state of the system is specified by the intensive
parameters of $\mathcal{B}$ (temperature, chemical potential etc.). The heat
evolved is transferred to the bath\cite{lie,dng,jpcm}; The motion of
system has the same frequency as the external field.

\subsection{Spatial coarse grained average}

\label{sta}

To describe the finite spatial resolution in a macroscopic measurement, for
each microscopic quantity $\xi^{\Psi^{\prime}}(\mathbf{r},t)$ defined in
pure state $\Psi^{\prime}(t)$, one introduces\cite{rus,bins} a truncated
quantity $\overline{\xi^{\Psi^{\prime}}}$ :%
\begin{equation}
\overline{\xi^{\Psi^{\prime}}}(\mathbf{R},t)=\int_{-\infty}^{\infty}d^{3}r%
\xi^{\Psi^{\prime}}(\mathbf{R}-\mathbf{r},t)f(\mathbf{r}),   \label{tru}
\end{equation}
where the scalar weight function $f(\mathbf{r})$ satisfies two conditions:
(i) $\int_{-\infty}^{\infty}d^{3}rf(\mathbf{r})=1$; and (ii) $F(\mathbf{k}%
)=\int_{-\infty}^{\infty}d^{3}re^{-i\mathbf{k}\cdot\mathbf{r}}f(\mathbf{r}%
)\rightarrow0$ for $k>k_{0}$. $k_{0}$ is solely determined by the type of
problem and calculation we have in mind\cite{bins}. Condition (ii) means
that the spatial Fourier components of the field variables are irrelevant
above a cut-off wave vector $k_{0}$. $\xi^{\Psi^{\prime}}$ is $%
\rho^{\Psi^{\prime}}$ or any Cartesian component of $\mathbf{e},$ $\mathbf{b}
$ and $\mathbf{j}^{\Psi^{\prime}}$. Using the truncated quantity $\overline{%
\xi^{\Psi^{\prime}}}(\mathbf{R},t)$ is stricter than the simple coarse
grained average $\Omega_{\mathbf{R}}^{-1}\int_{\mathbf{r}\in\Omega_{\mathbf{R%
}}}d^{3}r\xi^{\Psi^{\prime}}(\mathbf{r},t)$ for solids\cite{bins}, where $%
\Omega_{\mathbf{R}}$ is a physical infinitesimal volume around point $%
\mathbf{R}$.

The Maxwell equations for the truncated fields are%
\begin{equation}
\nabla _{\mathbf{R}}\cdot \overline{\mathbf{b}^{\Psi ^{\prime }}}=0,\text{ \
\ }\nabla _{\mathbf{R}}\times \overline{\mathbf{e}^{\Psi ^{\prime }}}%
=-\partial \overline{\mathbf{b}}/\partial t,  \label{tho}
\end{equation}%
and

\begin{equation}
\nabla _{\mathbf{R}}\cdot \overline{\mathbf{e}^{\Psi ^{\prime }}}=\overline{%
\rho ^{\Psi ^{\prime }}}/\epsilon _{0}\text{, \ \ }c^{2}\nabla _{\mathbf{R}%
}\times \overline{\mathbf{b}^{\Psi ^{\prime }}}=\overline{\mathbf{j}^{\Psi
^{\prime }}}/\epsilon _{0}+\partial \overline{\mathbf{e}^{\Psi ^{\prime }}}%
/\partial t.  \label{tinh}
\end{equation}%
The arguments of $\overline{\mathbf{e}^{\Psi ^{\prime }}}$, $\overline{%
\mathbf{b}^{\Psi ^{\prime }}}$, $\overline{\mathbf{j}^{\Psi ^{\prime }}}$
and $\overline{\rho ^{\Psi ^{\prime }}}$ are $(\mathbf{R},t)$, the spatial
differential $\nabla _{\mathbf{R}}$ is respect to the spatially coarse
gained coordinate $\mathbf{R}$. Because the spatial coarse graining (\ref%
{tru}) is a linear map, and\ eqs.(\ref{hom},\ref{inh}) are linear, eqs.(\ref{tho},\ref{tinh}) are also linear about
all truncated quantities appear in them.

If one distinguishes the free charges and bound charges in $\overline {%
\rho^{\Psi^{\prime}}}$ and $\overline{\mathbf{j}^{\Psi^{\prime}}}$, various
order electric and magnetic moments of the bound charges also appear as the
sources of fields\cite{rus,bins}. Such a description appears in
textbooks\cite{jac,pan}, and is convenient for the crystalline metals, alloys
(there is a clear distinction between free carriers and bound charges) and
insulators at low frequency. In amorphous semiconductors, the hopping
probability for localized carriers is\cite{epjb} $\sim10^{12}-10^{13}$sec$%
^{-1}$. On a time scale shorter than $10^{-12}$sec, one cannot distinguish a
localized carrier from a bound charge. Therefore for (i) metals, alloys and
insulators at high frequency (inter-band transition); and (ii) plasma\cite%
{pita} and amorphous semiconductors\cite{mic,long,dng,jpcm} at any
frequency, it is convenient not to distinguish the contributions from
carriers and from bound electrons in $\overline{\rho^{\Psi^{\prime}}}$ and $%
\overline{j_{\alpha}^{\Psi^{\prime}}}$, i.e. not to make the multipole
expansion.

For the following three reasons, we do not need\cite{rus,bins,jac} to take an
additional temporal coarse grained average over fields and sources at any
stage. First of all, eqs.(\ref{hom},\ref{inh}) are linear in all
quantities appearing in them: all the quantities are additive. Therefore for
typical cut-off wave vector $k_{0}=10^{6}$cm$^{-1}$ and number density of
particles $n=10^{22}$cm$^{-3}$, the relative fluctuation of a truncated
quantity is small\cite{jac} $\propto n^{-1/2}k_{0}^{3/2}\sim 10^{-2}$. The spatial average (\ref{tru}) is enough. Secondly, the time $%
v^{-1}k_{0}^{-1}$ spent by a particle with speed $v\sim10^{6}$m$\cdot $sec$%
^{-1}$ traversing a distance $k_{0}^{-1}$ is $10^{-14}$sec, is still in the
range of atomic or molecular motions\cite{jac}, averaging over such a time
period after spatial average (\ref{tru}) is pointless. Third, before the
spatial average (\ref{tru}), if one averages over a time period shorter than
the time scale of atomic and molecular motions, one cannot eliminate fast
fluctuation\cite{jac}. On the other hand, if one averages over a time period
longer than the time scale of atomic and molecular motions, one smeared or
even eliminates  the scattering phenomenon\cite{lv8}.

To obtain a macroscopic observable from the corresponding microscopic
quantity in the pure state, we first truncate the microscopic quantity in a
given pure state and then take an ensemble average. For a group of charged
particles interacting with an electromagnetic field, we can replace the
ensemble average with an average over all possible initial pure states.
Since for a given external field, the coupling between field and the system
can be included with additional terms [represented by $H_{fm}(t)$ in eq.(\ref%
{fm})] to the system Hamiltonian $H$, the state $\Psi ^{\prime }(t)$ of
system at time $t$ is completely determined by the value of $\Psi ^{\prime }$
at a previous moment through eq.(\ref{ts}). If we
adiabatically introduce an external field, then the ensemble average is changed
into an average over various initial values $\Psi ^{\prime }(-\infty )$ of
state $\Psi ^{\prime }$. Denote
\begin{equation}
\xi (\mathbf{R},t)=\sum_{\Psi ^{\prime }}W[\Psi ^{\prime }(-\infty )]%
\overline{\xi ^{\Psi ^{\prime }}}(\mathbf{R},t),  \label{mds}
\end{equation}%
where $\overline{\xi ^{\Psi ^{\prime }}}$ is any truncated quantity: $%
\overline{\rho ^{\Psi ^{\prime }}}$ or any Cartesian component of $\overline{%
\mathbf{e}^{\Psi ^{\prime }}}$, $\overline{\mathbf{b}^{\Psi ^{\prime }}}$
and $\overline{\mathbf{j}^{\Psi ^{\prime }}}$. $W[\Psi ^{\prime }(-\infty )]$
is the probability that the system is initially in state $\Psi ^{\prime
}(-\infty )$. $W[\Psi ^{\prime }(-\infty )]$ depends only on the energy of $%
\Psi ^{\prime }(-\infty )$, can be taken as either a canonical or a grand
canonical distribution (p678 of \cite{lie}). Then $\xi (\mathbf{R},t)$ is
the usual macroscopic observable\cite{dng}.

Because eqs.(\ref{tho},\ref{tinh}) are linear, we can average the fields and
sources over all possible initial pure states of the system. Then we obtain
the macroscopic Maxwell equations:%
\begin{equation}
\nabla \cdot \mathbf{B}=0,\text{ \ \ }\nabla \times \mathbf{E}=-\partial
\mathbf{B}/\partial t,  \label{mho}
\end{equation}%
and%
\begin{equation}
\nabla \cdot \mathbf{E}=\rho /\epsilon _{0}\text{, \ \ }c^{2}\nabla \times
\mathbf{B}=\mathbf{j}/\epsilon _{0}+\partial \mathbf{E}/\partial t.
\label{mih}
\end{equation}%
Here we used the familiar symbols $\mathbf{E}$ and $\mathbf{B}$ to denote
the ensemble average of $\overline{\mathbf{e}^{\Psi ^{\prime }}}$ and $%
\overline{\mathbf{b}^{\Psi ^{\prime }}}$. The arguments of $\mathbf{E}$, $%
\mathbf{B}$, $\mathbf{j}$ and $\rho $ are $(\mathbf{R},t)$, $\nabla $
represents the operator respect to $\mathbf{R}$. The macroscopic sources ($%
\rho $ and $\mathbf{j}$) and fields depend on the intensive parameters of a
chosen ensemble.

In view of facts that (i) the microscopic eqs.(\ref{hom},\ref{inh}) are
linear; and (ii) both the ensemble average and truncation (\ref{tru}) are
linear maps of the microscopic fields and sources, to obtain eqs.(\ref{mho},\ref{mih}),
we can actually first take the ensemble average over the microscopic quantities,
and then truncate the ensemble averaged quantities.

The derivation suggested in this section differs from previous ones%
\cite{maz,ram,gro,pita} in two aspects: (i) to describe the electromagnetic
phenomena in macroscopic media, temporal coarse graining is unnecessary%
\cite{jac}. Only a spatial coarse-grained and ensemble average are needed; (ii)
Because for a given external field, the interaction of the system and field can
be written with additional terms in the Hamiltonian of system\cite{kth}. Then
for each wave function $\Psi^{\prime}$\ which satisfies eq.(\ref{ts}), one
may introduce the microscopic response $\mathbf{j}^{\Psi^{\prime}}$ for that
state. Thus the ensemble average eq.(\ref{mds}) can be delayed to the final
stage\cite{mic,dng} rather than taking at the outset\cite{kth,maz,ram,gro}.

Since eqs.(\ref{hom},\ref{inh}) have the same structure as eqs.(\ref{mho},\ref{mih}), to obtain the radiation fields produced by a mixed state, we only need to replace $\partial\mathbf{j}^{\Psi^{\prime}}(\mathbf{x}^{\prime},t^{\prime})/\partial
t^{\prime}$ with $\partial\mathbf{j}(\mathbf{x}^{\prime},t^{\prime})/\partial
t^{\prime}$.

\subsection{Polarization, current density and the source of radiation field}

For a group of charged particles obeying CM, the induced dipole moment is
determined by the displacement of charges. The current density is
proportional to the velocities of charges. The radiation field is caused by
the accelerations of charges. We will show that  for a group of charged particles obeying QM, there exist similar relations among polarization, current density and the source of radiation field: the current density is time derivative of polarization, and the source of radiation field is time derivative of current density.

Eqs.(\ref{cur},\ref{era},\ref{scf}) have
shown that the time derivative relation is correct for the microscopic
quantities in a pure state.
It is easy to see that the time derivative relation is also correct for the
macroscopic quantities. For the truncated quantity $\overline{\mathbf{P}%
^{\Psi^{\prime}}}(\mathbf{R},t)=\int_{-\infty}^{\infty}d^{3}r\mathbf{P}%
^{\Psi^{\prime}}(\mathbf{R}-\mathbf{r},t)f(\mathbf{r})$ and the macroscopic
polarization $\mathbf{P}(\mathbf{R},t)=\sum_{\Psi^{\prime}}W[\Psi^{\prime
}(-\infty)]\overline{\mathbf{P}^{\Psi^{\prime}}}(\mathbf{R},t)$, eq.(\ref%
{cur}) implies%
\begin{equation}
\overline{j_{\alpha}^{\Psi^{\prime}}}(\mathbf{R},t)=\frac{\partial \overline{%
P_{\alpha}^{\Psi^{\prime}}}(\mathbf{R},t)}{\partial t}\text{ and\ }%
j_{\alpha}(\mathbf{R},t)=\frac{\partial P_{\alpha}(\mathbf{R},t)}{\partial t}%
.   \label{pcr}
\end{equation}
Sometimes, the second equation in (\ref{pcr}) is taken as a redefinition of
polarization\cite{pita,ald}. Since eqs.(\ref{mho},\ref{mih}) have the same
structure as eqs.(\ref{hom},\ref{inh}), the macroscopic radiation field is
determined by $\partial j_{\alpha}(\mathbf{R},t)/\partial t$.

\subsection{Current density as the response of system to an external field}

In Sec.\ref{sta}, we have seen that (1) if we average the motion of particles over the microscopic time scale $v^{-1}k_{0}^{-1}$, the averaged charge and current density still oscillate in frequency $k_{0}v$; (ii) if we
average fields and sources over longer time scale (e.g. several periods of external field), the scattering phenomena are smeared\cite{jac,lv8}.
Therefore we should not take temporal coarse-grained average even for describing the electromagnetic phenomena in a mixed state of a macroscopic system.

On the other
hand, the induced charge density $\rho$ and the current density $\mathbf{j}$
are also the macroscopic responses of the system to an external field.
Therefore both the procedure in Sec.\ref{sta} and the time derivative
relation requires that temporal coarse graining should not be taken in the
macroscopic charge density and current density either as the sources of
fields or as the responses to external fields.

The microscopic response method (MRM) used this fact as the
starting point to calculate conductivity and Hall mobility\cite{mic,dng,long}%
. The entropy production is reflected in the existence of a steady state of the
system, which is in good contact with a heat and material bath. Because the system is in a
good thermal and chemical contact with a bath in an external field,
after a short transient period, the system will eventually reach a steady
state which oscillates at the same frequency as the external field. In the
steady state, the parameters characterizing the ensemble do not change: the
evolved or absorbed energy are transferred to or taken from the bath.
For a classical charged oscillator undergoing forced oscillation, it is
well-known that if the input energy is completely dissipated, the system
will eventually reach a steady state. With perturbation theory, one can show
that a solid in good thermal contact with a bath will reach a steady
state in an external field\cite{kubo,jpcm}.

Temporal coarse graining is a key step in the kinetic descriptions of the
irreversibility\cite{lie,kth}. For the processes caused by mechanical
perturbations, temporal coarse grained average can be avoided. Faber\cite%
{fab} and Mott\cite{sm} conjectured an expression for the ac conductivity for
strong scattering based upon assumptions: (i) in the Kramers-Heisenberg dispersion relation for
Raman scattering, put the final state and initial state the same; (ii)
require the zero frequency limit consistent with the Greenwood's dc
conductivity formula\cite{gre}, this was realized by taking long time limit
(i.e. temporal coarse graining) in the contracted Kramers-Heisenberg
relation. If the second step was not used, the first step would end up with
the conductivity formula (expressed by the single electronic states) derived
by the MRM in which no temporal coarse graining is taken\cite%
{mic,long,dng,kubo}. For crystalline metals, such a formula was written down
without derivation, cf. eq.(13.37) of \cite{ash}.

\section{Summary}

\label{sum} According to the semi-classical radiation
theory, the radiation fields are determined by the time derivative of the
current density. For a pure state of a many-body system, a strict current
density expression has been obtained from the microscopic response method.
Thus we have established microscopic Maxwell equations for a pure state.
For a charged quantal particle, three radiation fields involve vector potential only. This is a new example demonstrating the observability  of vector potential. Five radiation fields are perpendicular the canonical momentum  of the particle. In a many-body system, one of its radiation fields is perpendicular to
$\mathbf{A}(\mathbf{x}_{j},t)\times
\lbrack\nabla\times(\nabla_{j}\Psi^{\prime})]$,
 which does not
exist for a single charged particle. We predict that this form of radiation can be detected.

We have extended Russakoff-Robinson's \textit{ansatz }to a group of charged
particles obeying QM to derive macroscopic Maxwell equations. Only the
spatial coarse graining is relevant. The charge density and current density
as sources or responses is well-defined for a pure state. Every macroscopic
quantity can be obtained from the corresponding microscopic quantity by
taking spatial coarse graining and ensemble average, cf. eqs.(\ref{tru},\ref%
{mds}). Temporal coarse graining is inadequate for the macroscopic fields,
sources and the responses of system to field. The consistency among the
induced displacement, the current and the acceleration requires that one
should not take temporal coarse graining in the current density (transport
coefficients). This agrees with the microscopic response method\cite%
{mic,dng,jpcm,long}. The entropy production is reflected by the existence of
a steady state of the system in a good thermal connection with a bath.
If temporal coarse graining is taken in the transport coefficients obtained
by the microscopic response method, one can recover the corresponding
results obtained by the kinetic approaches.

\begin{acknowledgement}
This work is supported by the Army Research Laboratory and Army Research Office under Grant No. W911NF1110358 and the NSF under Grant DMR 09-03225.
\end{acknowledgement}

\end{document}